\documentclass{emulateapj}
\usepackage[below]{placeins}
\usepackage{natbib}
\slugcomment{To be published by the Astrophysical Journal}

\def\gtorder{\mathrel{\raise.3ex\hbox{$>$}\mkern-14mu
    \lower0.6ex\hbox{$\sim$}}}
\def\ltorder{\mathrel{\raise.3ex\hbox{$<$}\mkern-14mu
    \lower0.6ex\hbox{$\sim$}}}
\def\hmpc{h^{-1} \mathrm{Mpc}}
\def\hkpc{h^{-1} \mathrm{kpc}}
\def\kmsmpc{\ {\rm km~s^{-1} Mpc^{-1}}}
\def\msun{M_\odot}
\def\hmsun{h^{-1} M_\odot}

\def \ie {{\it i.e.,} }
\def \eg{{\it e.g.,}}

\def \br{{\bf r}}

\newcommand {\nbody} {$N$-body}

\newcommand {\sbh} {SBH}

\newcommand {\lcdm} {$\Lambda$CDM}

\shorttitle{QSOs, the dark side}
\shortauthors{Romano-Diaz, Shlosman, Trenti and Hoffman}

\begin{document}

\title{The Dark Side of QSO Formation at High Redshifts}

\author{
Emilio Romano-Diaz and Isaac Shlosman}

\affil{Department of Physics and Astronomy
University of Kentucky
Lexington, KY 40506-0055, USA}

\and

\author{Michele Trenti}

\affil{Department of Astrophysical \& Planetary Sciences, CASA
University of Colorado,
Boulder, CO 80309, USA}

\and

\author{Yehuda Hoffman}

\affil{Racah Institute of Physics
Hebrew University
Jerusalem 91904, Israel}


\begin{abstract}
Observed high-redshift QSOs, at $z\sim 6$, may reside in massive dark 
matter (DM) halos of more than $10^{12}~\msun$ and are thus expected 
to be surrounded by 
overdense regions. In a series of 10 constrained simulations, we have
tested the environment of such QSOs. The usage of Constrained 
Realizations has enabled us to address the issue of cosmic variance
and to study the statistical properties of the QSO host halos.
Comparing the computed overdensities
with respect to the unconstrained simulations of regions empty of QSOs,
assuming there is no bias between the DM and baryon distributions,
and invoking an observationally-constrained duty-cycle for Lyman Break 
Galaxies, we have obtained the galaxy count number for the QSO
environment. We find that a clear discrepancy exists between the computed 
and observed galaxy counts in the \citet{kim09} samples. Our
simulations predict that on average eight $z \sim 6$ galaxies per QSO
field should have been observed, while \citet{kim09} detect on average 
four galaxies per QSO field compared to an average of three galaxies in 
a control 
sample (GOODS fields). While we cannot rule out a small number statistics 
for the observed fields to high confidence, the discrepancy suggests that 
galaxy formation in the QSO neighborhood proceeds differently than in the 
field. We also find that QSO halos are the most massive of the simulated 
volume at $z\sim 6$ but this is no longer true at
$z\sim 3$. This implies that QSO halos, even in the case they are
the most massive ones at high redshifts, do not evolve
into most massive galaxy clusters at $z=0$. 
\end{abstract}



\section{Introduction}
\label{sec:intro}

QSOs are among the most luminous objects in the universe and can be
detected at high redshifts, when the universe was still very young ---
so far up to $z\sim 6.42$ \citep[\eg][]{fan03,willott10}. The comoving
space density at the bright end of the luminosity function \citep[as
detected by the Sloan Digital Survey, SDSS][]{york00} is $\sim (2.2
\pm 0.73)~h^3\mathrm{Gpc}^{-3}$ at $z\sim 6$
\citep{fan04}. Observations of high-$z$ QSOs raise a number of
fundamental questions: where and how they formed, and what is their
relationship to the formation of first stars and galaxies in the
universe. High-$z$ QSOs are important cosmological probes for studying
the star formation history, metal enrichment, early galaxy formation,
growth of supermassive black holes (\sbh{}s), properties of the
interstellar and intergalactic matter, and the epoch of re-ionization
in the universe.

Rare, high-$z$ QSOs have been claimed to form in highly overdense
regions of the initial matter distribution. Their central \sbh{}s are
expected to have grown fast via mergers and/or accretion.  Observed
QSOs at $z\sim 6$ appear to host SBHs of $\sim 10^9 \msun$ possibly
residing within the most massive halos of a few $\times 10^{12}
\msun$. Numerical simulations estimate a similar (to high-$z$ QSOs)
comoving density $\sim$ 1~Gpc$^{-3}$ of these massive halos
\citep[\eg][]{millenium,sijacki09}. This similarity however, depends
on the so-called QSO duty cycle, or a fraction of the time the QSO
is actually active. If this duty cycle is less than unity, the QSOs
are less rare and correspondingly reside in the less massive halos
(e.g., Overzier et al. 2009). Additional arguments have been used in 
favor of high-$z$ QSOs residing in lower mass halos (e.g., Willott et 
al. 2005).

\citet{kaiser84} and \citet{efst88} have shown that the presence of
high peaks (rich clusters) in the primordial density field enhances
their correlation function. Compared to clusters, galaxies have a
smaller correlation length $\xi(r) = 1$ for $r_g \approx 4-7~\hmpc$
\citep{davis83}. Therefore, primordial galaxy-size high peaks could in
principle increase the clustering of galactic mass-scale halos in
their vicinity \citep{munoz08}.

Population of galaxies responsible for re-ionization of the Universe
is not yet found.  The epoch of re-ionization which extended from
$z\sim 15$ has apparently ended by $z\sim 6$, as indicated by
observations \citep[\eg][]{becker01,barkana02,cen02} and by numerical
modeling \citep[\eg][]{gnedin97,haiman03,wyithe03}. Because high
overdensities will accelerate the galaxy formation and evolution,
it is natural to look for these galaxies in the QSO neighborhoods.

A significant excess of sources compared to the density seen in 
GOODS\footnote{Great Observatory Origins Deep Survey \citep{goods}} has
been obtained from analyzing the QSO J1030+0524 field at $z\sim 6.28$
\citep{stiavelli05}. \citet{zheng06} also observed a significant
overdensity\footnote{As pointed out by Overzier (2009), this overdensity
is less significant, due to the under-estimating the contamination from
lower $z$ interlopers. This comment applies also to Stiavelli et al. 
(2005) and Kim et al. (2009)}
around the SDSS QSO J0836+0054 at $z\sim 6$. These
observations suggest that galaxy clustering might win over a
prospective negative feedback of QSO on its environment, and an excess
of galaxies is associated with the QSO. However, a sample of $i_{\rm
  775}$-dropout candidate galaxies identified in five fields of the
{\it Hubble Space Telescope (HST)} Advanced Camera for Surveys (ACS)
centered on SDSS QSOs at $z\sim 6$ \citep{kim09,maselli09} hint to a
more complex behavior. Two fields has been claimed overdense, two
underdense and an additional one at the average density of GOODS
(\citeauthor{kim09}). \citet{willott05} detected no overdensities around
SDSS QSOs, although their survey has been less sensitive than Kim's et
al., including J1030+0524. 

In this paper, we model the formation of massive {\it pure} DM halos
at $z\sim 6$ by means of Constrained and Unconstrained numerical 
simulations (see section~2) and analyze the
possible causes for the apparent discrepancy between simulations and
observations of high-$z$ QSO environments described by \citet{kim09}.
We choose $\sim 10^{12}~\msun$ halos collapsing
by $z\sim 6$ as both realistic and representative case.
For this we resort to a sample of 10 different QSOs
environments simulated at high-resolution. Our approach is different 
and yet complementary to the study by Overzier et al. (2009),
who have constructed the semi-analytical galaxy catalogues based on
the {\it Millennium} simulation (Springel et al. 2005), and hence 
could make use of the sub-grid baryonic physics. 
The {\it Millennium} simulation volume $500^3~\mathrm{h^{-3} Mpc^3}$ 
is smaller than the typical volume occupied by a bright $z\sim 6$ QSO, 
if the QSO duty cycle is about unity (see above). However, it
may provide a lower limit for what is expected for even more massive 
halos.

Within the present cosmological formation scenario, \lcdm, structure
evolves in a 'hierarchical' way, with the growth of small density
fluctuations amplified by gravity from an otherwise smooth density
field. Within such scenario small objects collapse first and
subsequently merge to form progressively larger and more massive
structures. Therefore, the formation of high$-z$ galaxies and QSOs
depends on the abundance of DM halos within a given volume.

Similarity between the cosmic star formation history
\citep[\eg][]{madau96,bunker04} and the evolution of QSO abundances
\citep[\eg][]{shaver96} suggests a possible link between galaxy
formation and \sbh{} growth. Evidence supporting this relation comes
from the several correlations measured locally, i.e., at very low $z$,
between the \sbh{} masses and global properties of the host's spheroid
components, such as their masses and luminosities \citep{marconi03},
light concentration \citep{graham01} and stellar velocity dispersions
\citep{ferrarese00,gebhardt00}.

The growth of \sbh{}s can be plausibly linked to galaxy formation
process. Metal-rich gas associated with QSOs at high $z$
\citep[\eg][]{barth03} provides evidence that they are located at the
centers of massive galaxies. It is likely, therefore, that high-$z$
QSOs highlight the location of some of the first perturbations that
became nonlinear \citep[\eg][]{trenti07}. Such objects reside in
over-dense regions which might evolve into massive clusters of
galaxies containing a population of large elliptical galaxies. At low
redshifts, the \sbh{}s of these elliptical galaxies might be dormant
\citep[\eg][]{magorrian98, li07}. If the \sbh{}s evolve via gas
accretion, their buildup can be regulated by a radiative and
mechanical feedback onto the infalling material. However, if the main
mode of the \sbh{} growth comes from mergers, the feedback is
irrelevant.

Observations so far appear to confirm the existence of some feedback
effect on the QSO environment, but the results are not decisive. 
The large emission of radiation associated with the QSOs activity,
might be sufficient to ionize the surrounding IGM and could even
photoevaporate the gas from the neighboring DM halos, preventing star
formation before the gas cools down \citep[\eg][]{shapiro01} and
suppressing galaxy formation in its vicinity. This could lead to a
deficiency of (proto)galaxies around the QSOs despite the DM halo
excess. On the other hand, the QSO activity could also lead to
positive feedback, enhancing the star formation process and therefore,
galaxy formation \citep[\eg][]{begelman89,rees89,phinney89}.

Finally, evidence for a positive feedback from the QSOs has been
observed at lower redshifts, e.g., for Lyman-break galaxies at $z\sim
3-3.5$ \citep{steidel03}, and narrowband Ly$\alpha$ selected samples
of radio galaxies at $2 < z < 5.2$ \citep{venemans03}.
\citet{stevens10} used mid-infrared imaging of five QSO fields and
found evidence that the high-$z$ QSOs are associated with a
substantially elevated level of star formation activity at $1.7 < z <
2.8$. Radio-quiet QSOs have shown a lower excess of star formation
over radio-loud AGN at these redshifts.

This paper is organized as follows: our methodology and numerics are
explained in section~2, results are provided in section~3, and
comparison between observations and numerical simulations is given in
section~4. Section~5 discusses our results, and conclusions are given
in the last section.  The revised method for Constrained Realizations
implemented here is described in the Appendix.


\section{Methodology}

High-$z$ QSOs are simulated by following a large cosmological volume
to accommodate the very low space density of such population. Such
simulations exhibit a large dynamic range to ensue the hierarchical
build up of the hosts \citep[\eg][]{millenium} and may include the gas
dynamics during mergers, star formation, etc. \citep[][]{li07}.

An alternative approach is the use of the Constrained Realizations
method \cite[hereafter CRs,][]{bert87,hr91,vdw96} in order to
prescribe the formation and collapse of a suitably massive DM halo at
any redshift in a given computational box, with the subsequent
addition of the baryonic component.  In this approach one can
concentrate on the particular region of interest without the need to
resort to large and expensive computations.  Following this path, we
adopt the CR method of \citet{hr91} and impose the necessary
constraints to seed DM halos of a few$\times 10^{12}\msun$ at $z\sim
6$, assuming that they harbor the observed high-$z$ QSOs.

\subsection{Initial Conditions}
\label{sec:ic}

Our main goal is to assess the effects of a QSO-host DM halo on its
immediate region at high$-z$. For this purpose we have created a suit
of ten CRs with a DM halo seed of $10^{12}\hmsun$ collapsing by $z\sim
6$, according to the top-hat model. As a control sample, we have
constructed in tandem a suit of ten Unconstrained Realizations (UCRs,
hereafter). The same random seeds and cosmology have been used for
each pair of CR -- UCR realizations. Although ten realizations are not
statistically significant, it is a large enough sample to overcome the
cosmic variance.

The CRs were constructed as follows.
We have designed a set of ten different experiments, to probe
different merging histories (\ie environments) of a $10^{12}\hmsun$
halo in an \lcdm ~cosmology. The cosmological parameters are those
from the 5 years WMAP data release \citep{wmap5}, \ie $\Omega_{\rm m}
= 0.279$, $\Omega_\Lambda = 0.721$, and $h = 0.701$, where $h$ is the
Hubble constant in units of $100 \kmsmpc$.  $\sigma_8=0.817$, the rms
linear mass fluctuation within a sphere of radius $8 \hmpc$
extrapolated to $z=0$, was used to normalize the initial linear power
spectrum. The constraints were imposed onto a grid of $256^3$ within a
cubic box of $20 ~\hmpc$.  In the appendix \ref{sec:a1} we provide
details on how the constraints have been imposed. All constraints were 
imposed at the center of the computational box.

\begin{figure*}
  \begin{center}
    \includegraphics[angle=0,scale=0.8]{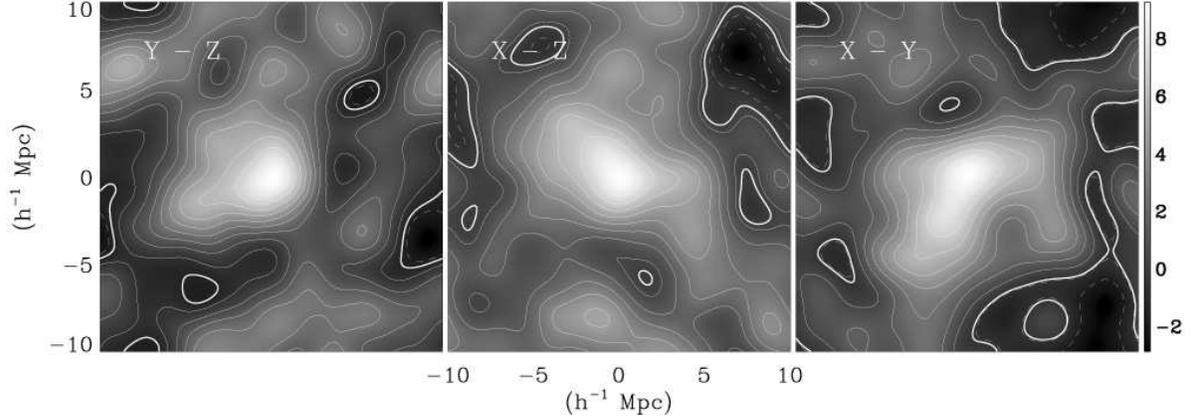}
  \end{center}
  \caption{Initial density field for one of the CR models. Each panel
    depicts a perpendicular cut of a density cut along the central
    slice. Continuous lines represent overdensity regions, dashed lines
    --- underdense regions, while the thick line is the mean overdensity
    within the box. The main constraint of $10^{12}\hmsun$ can be
    noticed at the center of each cut surrounded by the typical large scale
    structure of the Universe. The gray-scale vertical bar provides
    the linear density scale.}
\label{fig:ics}
\end{figure*}

Figure~\ref{fig:ics} shows the typical outcome of the CR procedure for
one of the models. Each panel shows a 2-D cut (perpendicular to each
other) along the central slice of the density field convolved with a
Gaussian filter of $10^{12}~\hmsun$. The continuous lines represent
overdense regions, the dashed lines --- the underdense ones, while the
thick continuous line is the average density within the box. The main
constraint can be clearly noticed as the central peak in each
panel. It is noteworthy to realize the presence of other smaller peaks
surrounding the constrained region. These peaks, which are due to the
random component of the CR method, could plausibly lead to major
mergers at later epochs with the central overdensity. Smooth
accretion of the surrounding matter, together with mergers, make the
halos grow continuously, by $z=3$ they have a mass $\sim 10^{13}\msun$
(see Sec~\ref{sec:mah}). Such halos, together with their surrounding
environment could potentially be the protocluster regions
\citep[\eg][]{venemas07} (see also Figure~\ref{fig:nbody}).

\begin{figure}
  \begin{center}
    \includegraphics[angle=0,scale=0.45]{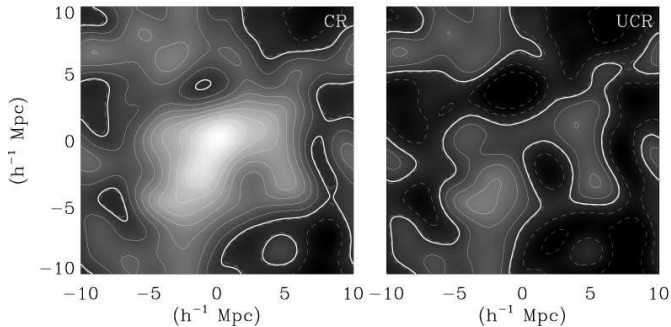}
  \end{center}
  \caption{2-D cuts of the central slice of thickness $78 ~\hkpc$, of
    the initial density fields for one of our CR models (left panel)
    and for its respective UCR counterpart (right panel). Despite the
    main feature missing in the UCR model, the overall LSS is the same
    in both models. The contour lines are the same for both models.}
\label{fig:ic-comp}
\end{figure}

The UCR set was constructed using the same cosmological parameters,
grid and box sizes and random seeds, without imposing any constraint
within the fields, resulting in a truly corresponding unconstrained
field of the CR one.  Figure~\ref{fig:ic-comp} depicts the central
slice of a CR model (left panel) and its UCR counterpart (right panel)
from our set. The normalization and contours for both fields are the
same in order to stress similarities and differences between the two
models. Clearly absent is the peak at the center. However, the overall
filamentary structure and underdense regions remain basically the same
in both models.

\subsection{Numerical simulations}
\label{sec:nbody}

The $10+10$ simulations were performed with vacuum boundary conditions
and physical coordinates. Due to these choices, a sphere of radius
$10~\hmpc$ was carved out from each of the initial fields and evolved
from $z = 199$ until $z=3$ by means of the FTM-4.5 code
\citep{hel94,heller07}. The mass resolution within our simulations,
under the cosmological model assumed, is of $m = 2.95\times 10^8
\msun$. Therefore, a galactic mass halo of $10^{11}\hmsun$ or above
can be resolved with at least 1,000 particles.

\begin{figure}
  \begin{center}
    \includegraphics[width=3.4in]{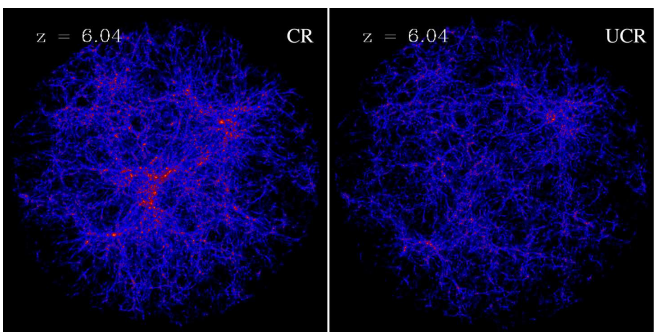}
  \end{center}
  \caption{Typical \nbody outcome at $z\sim6$ for a CR model (left
    panel) and its respective UCR (right panel). These models are the
    evolved fields of Fig.~\ref{fig:ic-comp}. Colors are proportional
    to their local densities. The box size is $20$~Mpc for both
    models.}
\label{fig:nbody}
\end{figure}

Figure~\ref{fig:nbody} depicts the typical outcome for two of our
simulations, one CR (left panel) and its corresponding UCR realization
(right panel) at $z\sim6$. Both models correspond to the initial
conditions maps shown in Fig.~\ref{fig:ic-comp}. Colors in both plots
are proportional to their local particle density and they have been
equally normalized for a better comparison. The box size in both cases
is $20$ Mpc. The most striking feature in the CR frame is the
clustering of several halos of similar sizes and masses (see
sec.\ref{sec:mf}) around the central peak. Such clustering is absent
in the UCR. One could naively assume that since there is only one
constraint imposed onto the CR model, there should be only one halo of
such mass present in the field. However, this is not the case as
depicted in Fig.~\ref{fig:nbody}, a situation present in all of our
models. The presence of other equally massive or slightly smaller
structures around the QSO-constrained halo are due to the fact that
the imposed constraint excites the formation of structures around
it. This provides a higher density environment with respect to a
``normal'' environment, as shown in the right panel of
Fig.~\ref{fig:nbody}. Such a behavior is a natural
consequence of the hierarchical nature of the underlying CDM model.

\subsection{Halos}
\label{sec:halos}

\begin{figure}
  \begin{center}
    \includegraphics[width=3.4in]{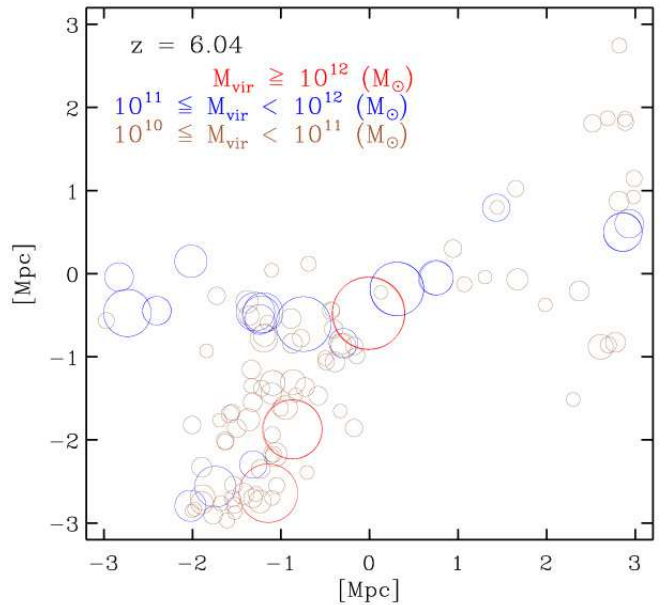}
  \end{center}
  \caption{Halo distribution at $z\sim6$ for a CR model (same as in
    previous figures). Circle sizes are proportional to their
    masses. The box size is 6~Mpc.}
\label{fig:halos}
\end{figure}

DM halos were identified by means of the HOP algorithm \citep{hop}.
This method isolates structures according to a purely density particle
criteria. Densities have been calculated locally using an Smooth
Particle Hydrodynamics (SPH) kernel.  A structure or halo was defined
as that region with a 3-D overdensity contour 200 times the mean
density.  The lower cut in halo mass resolution was at $N=100$
particles or $2.95\times 10^{10} \msun$ to ensure that the derived
halo masses are robust \citep{trenti10b}. Figure~\ref{fig:halos} shows
(central region of 6~Mpc) the typical outcome for the CR set. The
circles are proportional to their virial mass. Notice that apart from
the central halo with a mass $> 10^{12}\msun$, there are two other
halos of the same mass order.  Furthermore, the presence of several
$10^{11}$ and in particular of $10^{10}\msun$ halos is
overwhelming. The latter halos follow the distribution of the more
massive halos, indicating that the mass growth of such halos will
continue substantially. The general excess of halos at any mass scale
with respect to a ``normal'' (unconstrained) field can be easily
noticed in their respective mass functions.

\newpage
\section{Results}

\subsection{Halo mass function}
\label{sec:mf}

\begin{figure}
  \begin{center}
\includegraphics[width=3.65in]{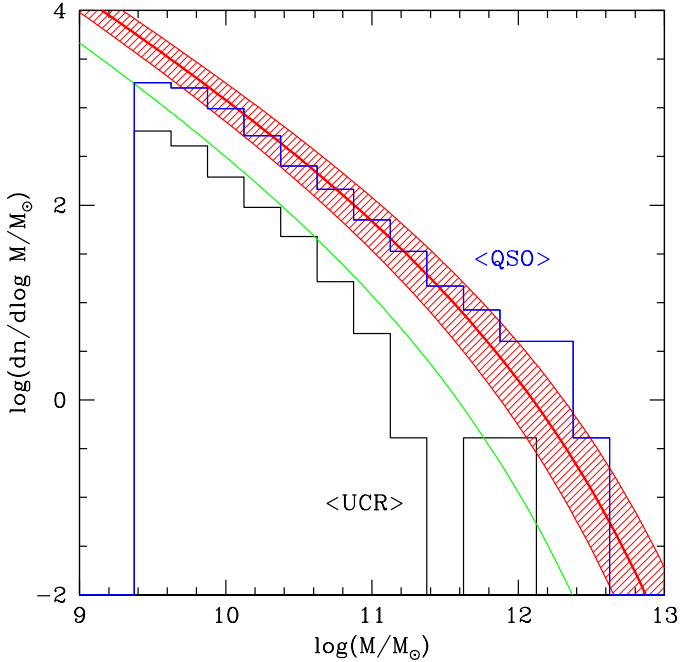}
  \end{center}
  \caption{Average Mass Functions for the two sample sets within a
    sphere of radius of $6~\hmpc$. The continuous lines are the
    expected number counts derived from the Extended Press-Schechter
    (EPS, Bond et al. 1991) formalism, for QSOs (red line) and UCRs
    (green line).  The shaded areas represent the 68\% confidence
    interval for a single realization of a CR run as derived from the
    linear overdensity variance among the sample of CR
    runs. Effectively, because we are showing a mass function averaged
    over 10 realizations, the shaded area is more significant.}
\label{fig:mf}
\end{figure}

\begin{figure*}
  \begin{center}
    \includegraphics[width=5.4in]{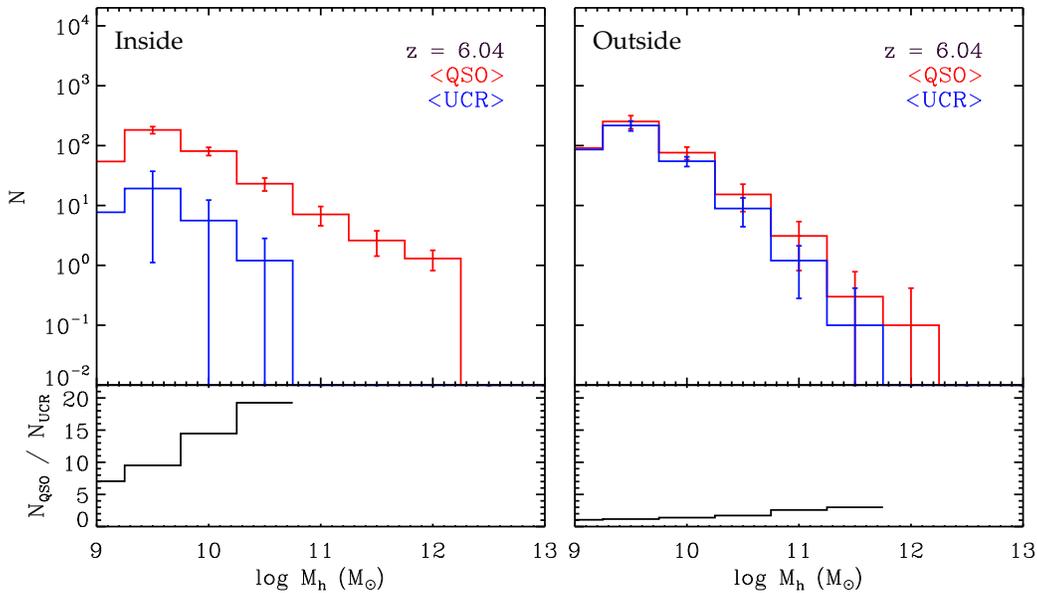}
  \end{center}
  \caption{Average Mass Function for the two sample sets within a box
    of 6~Mpc (left panel), and outside this inner box (right panel).}
\label{fig:mf6}
\end{figure*}

We have constructed halo Mass Functions (MF) from the halo catalogs
for each model set, CR and UCR respectively. Figure~\ref{fig:mf} shows
the total MFs averaged for the two sets of 10 realizations, and the
comparison Extended Press-Schechter curve \citep[EPS][]{bond91} in a
spherical subvolume of radius $6~{\rm Mpc}~h^{-1}$ around the QSO halo
location for each corresponding QSO -- UCR simulation pair in our
$10+10 set$.  Predictions for the mass function of the QSO halos runs
have been obtained by taking into account the average linear
overdensity inside this sphere introduced by imposing the constraint.
The presence of such overdensity changes the effective cosmology
\citep[\eg][]{goldberg04}. The shaded area represents the 68\%
confidence interval, derived from the $\pm 1 \sigma$ variation in the
linear overdensity distribution from the set of 10 CR runs. Noteworthy
is the effect of the constraints imposed on the field at all mass
scales.  The MFs show that there is not only a difference at the high
mass end range (imposed constraints regime), but the difference
persists over the entire mass range. This result is in agreement with
the hierarchical nature of the underlying CDM scenario. One would
expect a better correspondence between the two MFs in the lower mass
end, since the formation of smaller mass halos $(M < 10^{10}\msun)$ is
more common at these high redshifts. Indeed, such an agreement can be
better noticed in Figure~\ref{fig:mf6}, where we split the MFs
into two regions. The left panel represents the MFs of the inner box
of 6~Mpc side centered around the main constraint (in the UCR cases we
used the CR counterpart centers). Within this region the differences
between both curves should be more striking. Outside this region, the
corresponding MFs should have a better agreement as can be noticed on
the right panel of Fig.~\ref{fig:mf6}. The two MFs differ in less than
$5\%$ overall (lower panel), which can be attributed to the fact that
the region influence by the constraint is somewhat larger than a
radius of 3~Mpc.

The corresponding inner MFs show that the environment around a
prospective QSO-host halo should be very rich in (sub)structure, at
mass levels smaller than $10^{12}\msun$. A massive halo appears not
only to induce the formation of much smaller-scale, $M < 10^{10}\msun$
halos, which one would expect to find here anyway, but also on all
intermediate mass-scales which are rare at such redshifts. The
relatively high amount of $10^{11}\msun$ halos ($\sim 10$ on average)
exceeds the expectations of those from unconstrained simulations, as
shown by the total UCR MF (Figs.~\ref{fig:mf},~\ref{fig:mf6} right
panel).

Figures~\ref{fig:mf} and~\ref{fig:mf6} show that the potential QSO
fields reside in anomalously dense environments.


\subsection{MAH: Fate of the QSO-host halos}
\label{sec:mah}

The next logical step is to ask what is the fate of such high-density
regions? Do they evolve into the most massive clusters nowadays or to
a more ``normal'' cluster-group size halos? Majority of the performed
work in this field indeed assumes that high-redshift QSOs can exist
only in the precursors of what is now the most massive clusters ($m >
10^{15} \msun$) in the universe \cite[\eg][]{millenium,li07}.  Such an
assumption seems natural since the DM halos can only grow with time.

However, it has been shown that this rule might not
necessarily be true, both by means of the EPS formalism and by the
analysis of the {\it Millennium} simulation \citep{millenium} merger
tree history \citep{lucia07,trenti08b,overzier09}. The present DM 
mass of the QSO-host halos identified at
$z \sim 6$ could be of the order of $m \sim 10^{14} \msun$, and most
massive halos identified at various redshifts do not necessarily
maintain this property at lower redshifts.

This effect is confirmed in our simulations, although they have been
stopped at $z=3$. Throughout their evolution, the halos have
experienced a series of major mergers and have grown to $\sim 6 \times
10^{12}\msun$.  The most massive halo at $z=3$, the QSO9 has grown to
$\sim 10^{13}\msun$, although it is far from being most massive at
$z\sim 6$. We have constructed the Mass Accretion Histories (MAHs) for
each of the QSO-host halos in our CRs set. For this purpose, we have
reproduced their respective merger trees and followed the branch that
represents the constrained halo, at $z=6$, imposed by the initial
conditions.

\begin{figure}
  \begin{center}
    \includegraphics[width=3.4in]{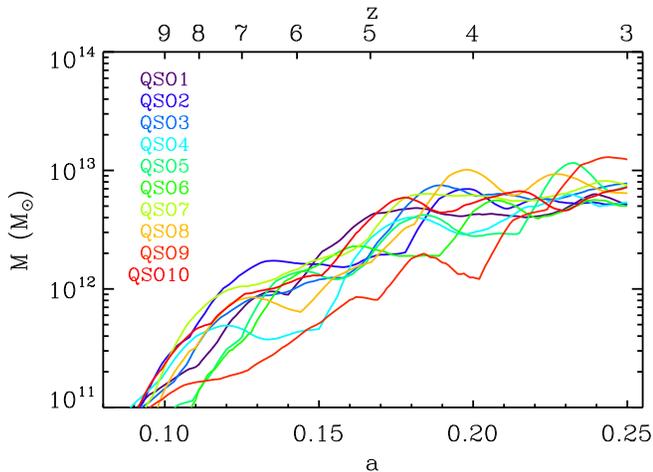}
  \end{center}
  \caption{Mass Accretion Histories (MAHs) for all CR models as a 
function of redshift $z$ (upper axis) and cosmological expansion
factor $a$ (lower axis).}
\label{fig:mah}
\end{figure}

Figure~\ref{fig:mah} represents the MAHs for all CR models.  Most of
them (apart from model QSO9) follow the similar trajectories. The
abrupt mass increases in the MAH curves are associated with major
mergers \cite[\eg][]{erd06,erd07}, which seem to occur at all
calculated redshifts.  If we compare our MAHs with the cluster sample
of \citet{t04} and their fitted MAH, we find that our halos will end
up as a normal cluster size halos. Furthermore, a simple estimate
accounting for all the mass that could collapse into such DM halos
from $z=3$ till the present time, indicates that halos will grow to
$\sim 9\times 10^{13} - 2 \times 10^{14} \msun$.

The natural dispersion of the MAH curves in Fig.~\ref{fig:mah} appears
to be the result of the cosmic variance and comes from varying the
seed of the initial conditions. A striking feature of these curves is
their relative evolution. For example, the curve corresponding to
QSO10 appears to be least massive of the sample at $z\sim 6$ while it
grows fast and becomes the most massive at $z\sim 3$.


\section{Comparison with observations: predictions for galaxy counts
in QSO fields}
\label{sec:observations}
 

\begin{deluxetable}{crrr}
\tablewidth{0pt}
\tablehead{
QSO Fields & S1 & S2 & S3 \\
}
\startdata
GOODS  & $8.08\pm 2.84$ & $3.95\pm 1.99$ & $2.96\pm 1.72$ \\
J1030$+$0524  & +$5.92\pm 4.70$ & +$4.05\pm 3.46$ & +$7.04\pm 3.60$ \\
J1049$+$4637  & -$0.08\pm 4.02$ & -$1.95\pm 2.81$ & +$1.04\pm 2.64$ \\
J1148$+$5251  & -$5.08\pm 4.02$ & -$1.95\pm 2.81$ & -$2.96\pm 3.60$ \\
J1306$+$0356  & -$7.08\pm 4.02$ & -$3.95\pm 2.81$ & -$1.96\pm 2.43$ \\
J1630$+$4012  & +$2.92\pm 4.37$ & +$4.05\pm 3.46$ & +$2.04\pm 2.82$ \\
Mean          & -$3.40\pm 1.89$ & +$0.05\pm 1.38$ & +$1.04\pm 1.37$  \\
\enddata

\tablecomments{\tabletypesize{\small} Number of $i_{\rm 775}$-dropouts
  and Poisson Error by S/N and Color Limit from \citet{kim09}: (1)
  observed QSO fields; (2) extra-galaxies in the S1 sample (i.e., the
  GOODS field subtracted from the actual galaxy count in Kim et
  al. [2009]); (3) extra-galaxies in the S2 sample; (4) extra-galaxies
  in the S3 sample. Except in the GOODS line, all numbers have GOODS
  fields subtracted. The S1 -- S3 samples are defined in the
  text. These numbers do not include the target QSO galaxies. The
  signs ``+'' mean overdensity and ``--'' mean underdensity with
  respect to the GOODS fields (first line). The `Mean' value for S3
  refers to mean galaxy counts in 5 observed fields (not to excess
  counts).
\label{table:datakim}}
\end{deluxetable}

\begin{deluxetable*}{ccccccc}
\tablewidth{0pt}
\tablehead{
\colhead{ID} & \colhead{Duty-Cycle} & \colhead{$M_{\rm min}$} & \colhead{UCR}
& \colhead{CR (QSO)} & \colhead{Extra-galaxies expected}\\
\colhead{(\citeauthor{kim09})} & $\epsilon_{\rm DC}$ & \colhead{$(10^{10}~\hmsun$)} & Halos & Halos & }
\startdata
S3  &  1.0  & 11.90  &  0.15 &  13.4  & 13 \\
S2  &  1.0  & 11.10  &  0.2  &  14.5  & 14 \\
S1  &  1.0  &  8.80  &  0.6  &  18.6  & 18 \\	  
\\
S3  &  0.5  & 10.60  &  0.3  &  15.2  &  7 \\
S2  &  0.5  &  8.80  &  0.4  &  18.7  &  9 \\
S1  &  0.5  &  6.38  &  1.2  &  28.1  & 13 \\
\\
{\bf S3} & {\bf 0.2}  & {\bf 6.76} & {\bf 0.7} & {\bf 26.8}  & {\bf 5} \\
S2  &  0.2  &  5.84  &  1.0  &  31.2  &  6 \\
S1  &  0.2  &  4.08  &  3.0  &  48.5  &  9 \\
\\
S3  &  0.1  &  4.66  &  1.5  &  40.9  &  4 \\
S2  &  0.1  &  4.11  &  2.0  &  48.1  &  5 \\
S1  &  0.1  &  2.83  &  6.1  &  69.7  &  6 \\
\enddata

\tablecomments{\tabletypesize{\small} Predicted galaxy counts: (1) the
  samples from \citet{kim09}; (2) the duty cycle; (3) minimum DM halo
  mass --- $M_{\rm min}$; (4) the UCR halos; (5) the QSO halos counts
  which represent the average number of halos in the respective sets
  more massive than $M_{\rm min}$; (6) the extra number of galaxies
  expected. All values have been averaged over 10 CRs and 10 UCRs.
\label{table:results}}
\end{deluxetable*}

Pure DM simulations can be used to predict the expected Lyman Break
Galaxies number counts in the \citet{kim09} observations by making
additional assumptions.  The QSO host galaxy is not counted in the
following statistics.  First, for simplicity, we assume that one
galaxy is found per one DM halo.  Second, based on the field of view
geometry of the \citet{kim09} observations, we compute the comoving
volume probed in an ACS field of view, 11.3~arcmin$^2$, i.e., $6\times
6~(\hmpc)^2$ at $z\sim 6$, for $i$-dropouts. Assuming a pencil beam
geometry and redshift uncertainty of $\Delta z = 1$ centered at $z=6$,
we obtain $6\times 6\times 317~(\hmpc)^3$ from the cosmic variance
calculator\footnote{\url{http://casa.colorado.edu/\lower3pt\hbox{\char126}trenti/CosmicVariance.html}}
of \citet{trenti08}. The effective volume for the search is however
smaller due to both incompleteness and because the faint $i$-dropout
sources cannot be detected if they are found in the proximity (or
behind) brighter foreground galaxies and stars. Based on artificial
source recovery simulations carried out to search for $z \gtrsim 5$
galaxies \citep[\eg][]{oesch07}, we estimate that the effective volume
is typically one half of the pencil beam volume. Hence, we assume that
each ACS field in the \citeauthor{kim09} observations has probed $\sim
5,706~(\hmpc)^3$.

Two values for S/N color limits of $i_{\rm 775}- z_{\rm 850}=1.3$ and
1.5 have been considered. Choice of S/N $> 5$ and $i_{\rm 775}-z_{\rm
  850} > 1.3$ have been labeled S1; S/N $> 5$ and $i_{\rm 775}-z_{\rm
  850} > 1.5$ have been labeled S2; and S/N $> 8$ and $i_{\rm
  775}-z_{\rm 850} > 1.3$ have been labeled
S3. 
\newpage

Table~\ref{table:datakim} reproduces the observational counts in
all QSO fields \citep{kim09} with one difference --- the GOODS counts
have been subtracted. So each number, with the exception of the GOODS
field, represents the extra-galaxy counts.

We use the galaxy counts in the GOODS field, as reported in the first
data line of Table~\ref{table:datakim}, to derive the number density
of $i$-dropouts galaxies in a typical region of the universe at $z
\sim 6$, depending on the different observation selection procedures
in \citeauthor{kim09}.  These number densities have been compared to
number densities of DM halos derived from our sample of unconstrained
control simulations.

The minimum DM halo mass, $M_{\rm min}$, required to match the
observed number density of galaxies has been calculated by assuming a
given duty cycle\footnote{The duty cycle is defined here as a
probability, $\epsilon_{\rm DC}$, ranging between 0 and 1, for a
given high-$z$ galaxy to be a Lyman Break Galaxy} $\epsilon_{\rm
  DC}$ for the Lyman Break Galaxies. The resulting $M_{\rm min}$ is
reported in Table~\ref{table:results} for different values of
$\epsilon_{\rm DC}$ and for the different selections S1 -- S3. The
values of $M_{\rm min}$ are consistent with those derived from a
conditional luminosity function approach
\citep{stark09,lee09,trenti10}, as well as from the clustering
properties of $i$-dropouts \citep{overzier06}.

Next, we have considered the CR runs and counted the average number of
halos with $M \geq M_{\rm min}$ in the proximity of a QSO host halo,
using a pencil beam volume of $6 \times 6 \times 12~(\hmpc)^3$
centered at the QSO position. The volume size is determined by the ACS
field of view and by the limits imposed by the size of our simulation
box. The depth of $12~\hmpc$ guarantees that the volume considered
does not extend outside the simulation computational volume, which is
progressively reduced as the overdensity imposed at the center of the
box collapses. We have subtracted the number of UCR halos above $M
\geq M_{\rm min}$ (column~4 of Table~\ref{table:results}) from that of
QSO halos in the pencil beam volume around the QSO (column~5 of
Table~\ref{table:results}). After multiplying this excess number by
$\epsilon_{\rm DC}$, we have obtained the excess number of galaxies
predicted in a QSO field compared to the GOODS field (last column in
Table~\ref{table:results}). For a typical duty cycle values of 0.2
\cite[\eg][]{wyithe06,stark07,lee09,trenti10}, we expect $\sim 5-9$
more galaxies in the QSO fields, depending on the observational
selection criteria, S1 -- S3. We note also that while the above value
of $\epsilon_{\rm DC}$ appears as a peak in the likelihood countours,
it is a very shallow maximum \cite[\eg][]{stark07}.  A variation in
$\epsilon_{\rm DC}$ value between 0.1 and 0.5 does not substantially
change the result of the galaxy counts within each of the samples, S1
-- S3.

The galaxy counts in Table~\ref{table:results} do differ when various
samples are compared. If we restrict ourselves to the conservative S3
sample, because it has the highest S/N value, the predicted
overdensity corresponds to $\sim 5$ extra galaxies, averaged over 
10 QSO fields.

Comparison between the observational counts in
Table~\ref{table:datakim} and predictions from numerical simulations
in Table~\ref{table:results} shows that a broad disagreement exists
between both counts of extra galaxies. Observational counts in each
sample are consistent with {\it no galaxy overdensity or underdensity
  in the QSO fields}, except perhaps for QSO J1030 field (overdensity)
and J1148 (underdensity). If we put more weight on S3, because it has
the highest S/N value, this conclusion strengthtens even more. The
average value of galaxy counts in QSO fields in excess of the GOODS
fields for the S3 selection is $1.04\pm 1.37$ which is much smaller
than the predicted 5 galaxies. This clearly suggests that galaxy
formation around a QSO halo does not follow the predictions based on
the abundance of DM halos in this environment compared to the
field. This discrepancy is further discussed in section~5. A similar
conclusion has been also reached by Overzier et al. (2009) based on
the counts of simulated dropout galaxies.

Finally, we note that model halos around the QSO are distributed
asymmetrically due to the presence of filaments converging at the QSO
location (see Figure~\ref{fig:halos}), hence the asymmetry of sources
observed in Fig.~6 of \citeauthor{kim09} is a natural consequence of
the topology of DM halo distribution around the QSO halo.


\section{Discussion}
\label{sec:discussion}

Observations by \citet{kim09} appear broadly consistent with there
being no overdensity or underdensity in the galaxy counts of QSO
fields with respect to the GOODS fields. This result is puzzling in
view of the substantial overdensity of DM halos in the vicinity of the
QSO host halo in the numerical simulations presented here.  Taken at
face value, the observations by Kim et al. imply that, at a given
mass, DM halos in the vicinity of a bright QSO host less luminous
galaxies than DM halos with the same mass in the control GOODS fields.

On the other hand, \citeauthor{kim09} find that the $i_{775}-z_{850}$
color distribution of dropout galaxies differs significantly from the
averages for the GOODS galaxies similarly selected, at 99\% confidence
level. This suggest that despite having a comparable average number of
galaxies, the QSO fields have observable differences compared to the
GOODS fields.

The simplest resolution of this discrepancy may lie in the relatively
low number statistics of the \citet{kim09} observations. Only five QSO
fields were observed and the number of galaxies detected per field is
low (3-8 on average depending on the selection adopted). In fact, if
the galaxy number counts for the QSO fields with S3 selection were
drawn from a Poisson distribution with $<N>=8$ (3 galaxies from GOODS
plus predicted excess of 5 galaxies), then there is a probability $p
\sim 0.04$ of measuring on average 4 galaxies per QSO field over the
five fields observed. Cosmic variance is likely to increase the field
to field variations, increasing $p$. But on the other hand, if the QSO
halo is more massive than $10^{12} M_{\sun}$ (as for example suggested
by \citealt{millenium}), $\sim 10^{13} M_{\sun}$ (e.g., Dijkstra et al.
2008), the excess number of galaxies in its surroundings will also be 
higher (see also Munoz \& Loeb 2008).

Of course, if the QSO halo is less
massive instead, then the \citeauthor{kim09} results are instead more
likely. As we stated in the Introduction, a number of arguments do
point to QSOs residing in the less massive halos (e.g., Walter at al. 
2004; Willott et al. 
2005; Overzier et al. 2009). We note, that although the duty cycles
of AGN, in general, and QSOs, in particular, are hotly debated issues at
present, the longest duty cycle among AGN appears to be $\sim 10^8$~yrs
--- that of radio galaxies. This hints at the high-$z$ QSO duty cycle
being less than unity. 

It is possible that the {\it projected} (on the sky) overdensities 
around the high-$z$ 
QSOs are not associated with the {\it actual} 3-D overdensities, and 
are the result of contamination by the `field' dropout galaxies (e.g., 
Overzier et al. [2009]). In their Fig.~15, Overzier et al. exhibit 
$\sim 30$ projected overdensities from the Mock Survey of 
$\sim 70,000$~arc~min$^2$ area based on the Millennium simulation. Many
of these overdensities do not correspond to the 3-D overdensities.
Therefore, we estimate the probability of such contamination in the
Kim et al. (2009) fields. Kim et al. used the ACS fields which correspond
to 1/30 of a 316~arc~min$^2$ GOODS field. Hence, each Kim et al. field
is $\sim 316\times 30 = 9,480$ times smaller than the Overzier et al. field.
Assuming here the worst case scenario that {\it all} of the 30 
overdense regions in their Fig.~15 are `fake' projection cases, we
obtain the probability of such a fake overdensity in Kim et al.
field as $30:9,480 = 0.0045$. For 5 fields considered by Kim et al.,
the combined probability for such a fake overdensity not to be 
associated with a QSO is $\sim 0.022$, and hence can be fully neglected
by us. Of course they are substantial over a much larger area considered
by Overzier et al. 

Alternatively to the low number statistics case, we should take into 
account that Kim et al. found one
$2\sigma$-$3\sigma$ overdensity, one --- a $2\sigma$ underdensity
field, and three additional ones, with no difference to the GOODS
fields (using highest S/N column in Table~1). If taken at face value
(that is neglecting likely small number statistics), their
observational results are consistent with DM halo clustering (i.e.,
high density regions) and an overpopulation of galaxies, on one hand,
and with the scenario in which additional effects (i.e., ionization or
hydrodynamics) could play a more dominant role. Accepting results of
pure DM simulations which show a clear overdensity of halos in the QSO
field, the problem can lie with our understanding of various feedback
mechanisms based on evolution of starbursts and the central SBHs,
i.e., in the plausible competition between the positive and negative
feedbacks which follow from a number of physical processes.  This
conclusion is weakened by the apparent coincidence between the three
QSO fields and the GOODS fields (Table~\ref{table:datakim}).

The fact that some of the observed QSOs are located in average density
regions does not necessary imply that they are also average DM
density regions. For example, they could be surrounded by baryon-poor
DM halos. A variety of negative feedbacks can contribute in removing
baryons from DM halos or halting the star formation at those
redshifts, such as the UV background radiation and re-ionization
(depending on when it occurred), tidal and ram-pressure (ablation)
stripping, and dynamical friction could play a role. On the other
hand, highly and poorly collimated outflows from QSOs can provide a
positive feedback and even trigger the star formation. The inclusion
of baryons in DM simulations can affect the properties of DM
structures \cite[\eg][]{erd09,erd-sbh,duffy10}.  High resolution
numerical simulations will be necessary to investigate these and other
effects.

Do the \citeauthor{kim09} density estimates reflect the actual local
densities or ``projected'' density enhancements along the pencil
surveys? Observational evidence from other QSOs at lower redshifts 
seems to confirm the Kim et al. conclusions --- the QSOs reside in
different environments \citep{stevens10}. In fact, Kim et al.
reports a confidence estimate at the $95\%$ level. \citet{maselli09},
using a different method than Kim et al., have arrived  at similar 
conclusions. Of course, it is not straightforward to relate the
mass accumulation and environment evolution of high-$z$ QSOs
with those at low redshifts.

In a recent development, Utsumi et al. (2010) have used the large
field of view of the Suprime-Cam to image of the most distant QSO
J2329-0301 at $z=6.42$ and an empty field for comparison. This field of
view is about two orders of magnitude larger than the
HST/ACS field of view we have considered here (approximately
$0.25~\mathrm{deg^2}$ versus approximately
$11~\mathrm{arcmin^2}$). Seven LBG candidates have been found in the QSO field
compared to only one LBG in the comparison field. The authors point
out that the statistical significance of this apparent overdensity is
difficult to estimate. Some evidence exists that the LBGs near the QSO
avoid the 3~Mpc region centered on the QSO --- if confirmed this hints
to a substantial feedback from the QSO. Extension of this study to 
additional
$z\sim 6$ QSOs would be very useful to confirm widespread presence of
the galaxy overdensity over large scale. After all, the first report
on the HST/ACS imaging campaign of $z\sim 6$ QSOs showed a clear
overdensity of LBGs for the single field analyzed (Stiavelli et
al. 2005).

We now return to the more pessimistic conclusion from our comparison
between the observational galaxy counts and their numerical
predictions from our simulations. Can this discrepancy be bridged if
one assumes bias between the DM and baryon distribution? Our
assumption that there is only one galaxy per DM halo could indeed be
an oversimplification.  However, an increase in the number of galaxies
per host halo will only aggravate the problem, as this will increase
the numerical counts.

An interesting by-product of our analysis of the MAH curves for the
QSO halos is the observation that the most massive halos at high
redshifts do not necessarily remain the most massive ones at lower
redshifts, as exhibited by Fig.~\ref{fig:mah} (see also De Lucia et
al 2007; Trenti et al. 2008; Overzier et al. 2009). There are 
corollaries
when looking for possible low-$z$ counterparts of high-$z$
objects. There are a number of issues related to this problem. First,
will the \sbh{} grow in tandem with its host DM halo?  If yes, we
expect \sbh{} masses of the order of a few$\times 10^{10}~\msun$ in
the contemporary universe. Using the comoving volume density of
high-$z$ QSOs we estimate at least one such ultra-massive \sbh{}
within a sphere of $z\ltorder 0.2$. There are no observational
constraints at present to rule out such possibility.

However, the situation is aggravated because the most massive halos
today have not been most massive at $z\sim 6$, as our simulations show
(see also \citealt{trenti08b} for an analysis of the Millennium
simulation merger tree history and EPS modeling). In turn, this
implies that in the local universe there are \sbh{}s more massive than
the descendants of the $z \sim 6$ QSOs, if the \sbh{} mass is
correlated with the DM halo mass \citep{el-zant03,ferrarese05}.
 That may be contradicted by observations as such massive
objects will profoundly change the galactic dynamics because the
radius of influence of a few$\times 10^{10}~\msun$ \sbh{}s will be of
the order of the visible galaxy.


\section{Conclusions}
\label{sec:conclusions}

We have performed a set of carefully designed DM \nbody{} simulations
aimed at studying the environment of QSO-host halos at $z\sim 6$.  A
set of 10 Constrained Realizations has been constructed, seeding them
with a $10^{12}\msun$ DM halo designed to collapse by $z\sim 6$ in the
top-hat scenario. As a control sample, we performed a set of 10
simulations that represent the Unconstrained Realizations of the same
CR set. We have constructed the halo Mass Functions from each
simulation and showed how on average, the QSO (i.e., constrained)
sample enhanced the DM halo formation in its vicinity due to pure
gravitational effects, when compared with its respective UCR sample.
Assuming that there is no bias in baryon distribution with respect to
the DM in our simulations and that the QSO duty cycle is unity, we 
have calculated the expected Lyman Break
Galaxies number counts and compared this against the observed QSO
fields of \citet{kim09}.

Our main result is that the pure DM numerical simulations predict a
strong overdensity around the QSO peaks. The observations of
\citet{kim09} either do not support this or provide a mixed result
with one possibly overdense field, one weakly underdense, and three
having the density of the GOODS field. The explanation for this
discrepancy can lie in the small number of QSO fields observed coupled
with the small number of galaxies detected per field. To rule out 
such possibility it would be important to extend the Kim et al. study 
to a larger number of $z \sim 6$ QSO fields.

If, however, the over/underdensity result will be confirmed by higher
S/N observations, the resolution is probably related to the complex
physics of feedback mechanism involving the QSO and possibly the
starbursts in the surrounding galaxies, or both.  With the present
analysis we are not capable (nor it was our intention) to establish
the role of feedback in the QSO-host galaxy formation
evolution. However, our results suggest that in the case of high
density QSO environments, the absence of a comparable enhancement of
galaxy counts or their lack would point to a strong radiative and/or
mechanical negative feedback from the QSOs, resulting in a strong
biasing between the DM and baryon distribution.

A simple analysis of the halo-growth in our CR sample together with
their MAHs, indicates that these halos will possess, by $z=0$, a mass
in the range of $\sim 10^{14} \msun$.  This is consistent with our
choice of a perturbation mass at the initial conditions.

The underdense field J1148 of \citeauthor{kim09} that is especially at
odds with our numerical expectations, opens the possibility to future
investigations adding baryons to simulations to explore the roles of
feedback in the evolution of QSO environment, star formation, among
other relevant physical phenomena.

But perhaps, not all high-$z$ QSOs are formed in high density peaks
but in less massive (and more common) halos of $\sim
10^{11}~\msun$. The MAHs of our models show that there is natural
dispersion (cosmic variance) for the halo masses at each $z$ --- halos
which are most massive at $z\sim 6$ in fact appear least massive at
$z\sim 3$. Hence, QSOs could be surrounded by a more ``typical''
environment, like in the case of J1048+4637 of the \citeauthor{kim09}
sample (see a similar conclusion by De Lucia \& Blaizot (2007) and
Overzier et al. [2009]). If this is the case, a scenario in which 
negative feedback dominates might not be the only solution. Alternatively,
$z\sim6$ QSOs residing in lower mass halos will require evolution
of the $M_\bullet-\sigma$ correlation with $z$.


\acknowledgments 
We are grateful to our colleagues, too numerous to list, for
discussions on various topics addressed here. I.S. thanks Tomotsugu 
Goto for helpful comments on Suprime-Cam observations of the QSO 
J2329-0301 field. I.S. acknowledges the
hospitality of JILA Visiting Fellows program and of the Copernicus 
Astronomical Center, Warsaw. This research has been
partially supported by NASA/LTSA/ATP/KSGC and the NSF grants to
I.S., by a grant from the ISF (13/08) to Y.H., and by the University of
Colorado Astrophysical Theory Program through grants from NASA and the
NSF to M.T.  All simulations were run using the BCX IBM Cluster at the
University of Kentucky Supercomputer Center.


\bibliography{bibliography}

\begin{thebibliography}{70}
\expandafter\ifx\csname natexlab\endcsname\relax\def\natexlab#1{#1}\fi

\bibitem[{{Bardeen} {et~al.}(1986){Bardeen}, {Bond}, {Kaiser}, \&
  {Szalay}}]{bbks}
{Bardeen}, J.~M., {Bond}, J.~R., {Kaiser}, N., \& {Szalay}, A.~S. 1986, \apj,
  304, 15

\bibitem[{{Barkana}(2002)}]{barkana02}
{Barkana}, R. 2002, New Astronomy, 7, 85

\bibitem[{{Barth} {et~al.}(2003){Barth}, {Martini}, {Nelson}, \&
  {Ho}}]{barth03}
{Barth}, A.~J., {Martini}, P., {Nelson}, C.~H., \& {Ho}, L.~C. 2003, \apjl,
  594, L95

\bibitem[{{Becker} {et~al.}(2001){Becker}, {Martini}, et al. }]{becker01}
{Becker}, R.~H., et al. 2001, \aj, 122, 2850

\bibitem[{{Begelman} \& {Cioffi}(1989)}]{begelman89}
{Begelman}, M.~C. \& {Cioffi}, D.~F. 1989, \apjl, 345, L21

\bibitem[{{Bertschinger}(1987)}]{bert87}
{Bertschinger}, E. 1987, \apjl, 323, L103

\bibitem[{{Bond} {et~al.}(1991){Bond}, {Cole}, {Efstathiou}, \&
  {Kaiser}}]{bond91}
{Bond}, J.~R., {Cole}, S., {Efstathiou}, G., \& {Kaiser}, N. 1991, \apj, 379,
  440

\bibitem[{{Bunker} {et~al.}(2004){Bunker}, {Stanway}, {Ellis}, \&
  {McMahon}}]{bunker04}
{Bunker}, A.~J., {Stanway}, E.~R., {Ellis}, R.~S., \& {McMahon}, R.~G. 2004,
  \mnras, 355, 374

\bibitem[{{Cen} \& {McDonald}(2002)}]{cen02}
{Cen}, R. \& {McDonald}, P. 2002, \apj, 570, 457

\bibitem[{{Davis} \& {Peebles}(1983)}]{davis83}
{Davis}, M. \& {Peebles}, P.~J.~E. 1983, \apj, 267, 465

\bibitem[{{De Lucia} \& {Blaizot}(2007)}]{lucia07}
{De Lucia}, G. \& {Blaizot}, J. 2007, \mnras, 375, 2

\bibitem[{{Dijkstra} {et~al.}(2008)}]{dijkstra08}
{Dijkstra}, M., {Haiman}, Z., {Mesinger}, A. \& {Wyithe}, J. 2008, \mnras, 391, 1961

\bibitem[{{Duffy} {et~al.}(2010){Duffy}, {Schaye}, {Kay}, {Dalla Vecchia},
  {Battye}, \& {Booth}}]{duffy10}
{Duffy}, A.~R., {Schaye}, J., {Kay}, S.~T., {Dalla Vecchia}, C., {Battye},
  R.~A., \& {Booth}, C.~M. 2010, \mnras, 405, 2161

\bibitem[{{Dunkley} {et~al.}(2009){Dunkley}, {Komatsu}, {Nolta}, {Spergel},
  {Larson}, {Hinshaw}, {Page}, {Bennett}, {Gold}, {Jarosik}, {Weiland},
  {Halpern}, {Hill}, {Kogut}, {Limon}, {Meyer}, {Tucker}, {Wollack}, \&
  {Wright}}]{wmap5}
{Dunkley}, J., et al. 2009, \apjs, 180, 306

\bibitem[{{Efstathiou} \& {Rees}(1988)}]{efst88}
{Efstathiou}, G. \& {Rees}, M.~J. 1988, \mnras, 230, 5P

\bibitem[{{Eisenstein} \& {Hut}(1998)}]{hop}
{Eisenstein}, D.~J. \& {Hut}, P. 1998, \apj, 498, 137

\bibitem[{{El-Zant} {et~al.}(2003){El-Zant}, {Shlosman}, {Begelman}, \&
  {Frank}}]{el-zant03}
{El-Zant}, A.~A., {Shlosman}, I., {Begelman}, M.~C., \& {Frank}, J. 2003, \apj,
  590, 641

\bibitem[{{Fan} {et~al.}(2004){Fan}, {Hennawi}, {Richards}, {Strauss},
  {Schneider}, {Donley}, {Young}, {Annis}, {Lin}, {Lampeitl}, {Lupton}, {Gunn},
  {Knapp}, {Brandt}, {Anderson}, {Bahcall}, {Brinkmann}, {Brunner}, {Fukugita},
  {Szalay}, {Szokoly}, \& {York}}]{fan04}
{Fan}, X., et al. 2004, \aj, 128, 515

\bibitem[{{Fan} {et~al.}(2003){Fan}, {Strauss}, {Schneider}, {Becker}, {White},
  {Haiman}, {Gregg}, {Pentericci}, {Grebel}, {Narayanan}, {Loh}, {Richards},
  {Gunn}, {Lupton}, {Knapp}, {Ivezi{\'c}}, {Brandt}, {Collinge}, {Hao},
  {Harbeck}, {Prada}, {Schaye}, {Strateva}, {Zakamska}, {Anderson},
  {Brinkmann}, {Bahcall}, {Lamb}, {Okamura}, {Szalay}, \& {York}}]{fan03}
{Fan}, X., et al. 2003, \aj, 125, 1649

\bibitem[{{Ferrarese} \& {Ford}(2005)}]{ferrarese05}
{Ferrarese}, L. \& {Ford}, H. 2005, Space Science Reviews, 116, 523

\bibitem[{{Ferrarese} \& {Merritt}(2000)}]{ferrarese00}
{Ferrarese}, L. \& {Merritt}, D. 2000, \apjl, 539, L9

\bibitem[{{Gebhardt} {et~al.}(2000){Gebhardt}, {Bender}, {Bower}, {Dressler},
  {Faber}, {Filippenko}, {Green}, {Grillmair}, {Ho}, {Kormendy}, {Lauer},
  {Magorrian}, {Pinkney}, {Richstone}, \& {Tremaine}}]{gebhardt00}
{Gebhardt}, K., et al. 2000, \apjl, 539, L13

\bibitem[{{Giavalisco} {et~al.}(2004){Giavalisco}, {Ferguson}, {Koekemoer},
  {Dickinson}, {Alexander}, {Bauer}, {Bergeron}, {Biagetti}, {Brandt},
  {Casertano}, {Cesarsky}, {Chatzichristou}, {Conselice}, {Cristiani}, {Da
  Costa}, {Dahlen}, {de Mello}, {Eisenhardt}, {Erben}, {Fall}, {Fassnacht},
  {Fosbury}, {Fruchter}, {Gardner}, {Grogin}, {Hook}, {Hornschemeier}, {Idzi},
  {Jogee}, {Kretchmer}, {Laidler}, {Lee}, {Livio}, {Lucas}, {Madau},
  {Mobasher}, {Moustakas}, {Nonino}, {Padovani}, {Papovich}, {Park},
  {Ravindranath}, {Renzini}, {Richardson}, {Riess}, {Rosati}, {Schirmer},
  {Schreier}, {Somerville}, {Spinrad}, {Stern}, {Stiavelli}, {Strolger},
  {Urry}, {Vandame}, {Williams}, \& {Wolf}}]{goods}
{Giavalisco}, M., et al. 2004, \apjl, 600, L93

\bibitem[{{Gnedin} \& {Ostriker}(1997)}]{gnedin97}
{Gnedin}, N.~Y. \& {Ostriker}, J.~P. 1997, \apj, 486, 581

\bibitem[{{Goldberg} \& {Vogeley}(2004)}]{goldberg04}
{Goldberg}, D.~M. \& {Vogeley}, M.~S. 2004, \apj, 605, 1

\bibitem[{{Graham} {et~al.}(2001){Graham}, {Erwin}, {Caon}, \&
  {Trujillo}}]{graham01}
{Graham}, A.~W., {Erwin}, P., {Caon}, N., \& {Trujillo}, I. 2001, \apjl, 563,
  L11

\bibitem[{{Haiman} \& {Holder}(2003)}]{haiman03}
{Haiman}, Z. \& {Holder}, G.~P. 2003, \apj, 595, 1

\bibitem[{{Heller} \& {Shlosman}(1994)}]{hel94}
{Heller}, C.~H. \& {Shlosman}, I. 1994, \apj, 424, 84

\bibitem[{{Heller} {et~al.}(2007){Heller}, {Shlosman}, \&
  {Athanassoula}}]{heller07}
{Heller}, C.~H., {Shlosman}, I., \& {Athanassoula}, E. 2007, \apj, 671, 226

\bibitem[{{Hoffman}(2009)}]{hoffman09}
{Hoffman}, Y. 2009, in Lecture Notes in Physics, Berlin Springer Verlag, Vol.
  665, Data Analysis in Cosmology, ed. {V.~J.~Martinez, E.~Saar,
  E.~M.~Gonzales, \& M.~J.~Pons-Borderia}, 565--+

\bibitem[{{Hoffman} \& {Ribak}(1991)}]{hr91}
{Hoffman}, Y. \& {Ribak}, E. 1991, \apjl, 380, L5

\bibitem[{{Kaiser}(1984)}]{kaiser84}
{Kaiser}, N. 1984, \apjl, 284, L9

\bibitem[{{Kim} {et~al.}(2009){Kim}, {Stiavelli}, {Trenti}, {Pavlovsky},
  {Djorgovski}, {Scarlata}, {Stern}, {Mahabal}, {Thompson}, {Dickinson},
  {Panagia}, \& {Meylan}}]{kim09}
{Kim}, S., {Stiavelli}, M., {Trenti}, M., {Pavlovsky}, C.~M., {Djorgovski},
  S.~G., {Scarlata}, C., {Stern}, D., {Mahabal}, A., {Thompson}, D.,
  {Dickinson}, M., {Panagia}, N., \& {Meylan}, G. 2009, \apj, 695, 809

\bibitem[{{Lee} {et~al.}(2009){Lee}, {Giavalisco}, {Conroy}, {Wechsler},
  {Ferguson}, {Somerville}, {Dickinson}, \& {Urry}}]{lee09}
{Lee}, K., et al. 2009, \apj,
  695, 368

\bibitem[{{Li} {et~al.}(2007){Li}, {Hernquist}, {Robertson}, {Cox}, {Hopkins},
  {Springel}, {Gao}, {Di Matteo}, {Zentner}, {Jenkins}, \& {Yoshida}}]{li07}
{Li}, Y., et al. 2007, \apj, 665, 187

\bibitem[{{{\L}okas} {et~al.}(2004){{\L}okas}, {Bode}, \& {Hoffman}}]{lokas04}
{{\L}okas}, E.~L., {Bode}, P., \& {Hoffman}, Y. 2004, \mnras, 349, 595

\bibitem[{{Madau} {et~al.}(1996){Madau}, {Ferguson}, {Dickinson}, {Giavalisco},
  {Steidel}, \& {Fruchter}}]{madau96}
{Madau}, P., {Ferguson}, H.~C., {Dickinson}, M.~E., {Giavalisco}, M.,
  {Steidel}, C.~C., \& {Fruchter}, A. 1996, \mnras, 283, 1388

\bibitem[{{Magorrian} {et~al.}(1998){Magorrian}, {Tremaine}, {Richstone},
  {Bender}, {Bower}, {Dressler}, {Faber}, {Gebhardt}, {Green}, {Grillmair},
  {Kormendy}, \& {Lauer}}]{magorrian98}
{Magorrian}, J., et al. 1998, \aj, 115, 2285

\bibitem[{{Marconi} \& {Hunt}(2003)}]{marconi03}
{Marconi}, A. \& {Hunt}, L.~K. 2003, \apjl, 589, L21

\bibitem[{{Maselli} {et~al.}(2009){Maselli}, {Ferrara}, \&
  {Gallerani}}]{maselli09}
{Maselli}, A., {Ferrara}, A., \& {Gallerani}, S. 2009, \mnras, 395, 1925

\bibitem[{{Mu{\~n}oz} \& {Loeb}(2008)}]{munoz08}
{Mu{\~n}oz}, J.~A. \& {Loeb}, A. 2008, \mnras, 385, 2175

\bibitem[{{Oesch} {et~al.}(2007){Oesch}, {Stiavelli}, {Carollo}, {Bergeron},
  {Koekemoer}, {Lucas}, {Pavlovsky}, {Trenti}, {Lilly}, {Beckwith}, {Dahlen},
  {Ferguson}, {Gardner}, {Lacey}, {Mobasher}, {Panagia}, \& {Rix}}]{oesch07}
{Oesch}, P.~A., et al. 2007, \apj, 671, 1212

\bibitem[{{Overzier} {et~al.}(2009){Overzier}, {Guo}, {Kauffmann}, {De Lucia},
{Bouwens} \& {Lemson}}]{overzier09}
{Overzier}, R.~A., {Guo}, Q., {Kauffmann}, G., {De Lucia}, G., {Bouwens}, R., 
\& {Lemson}, G.\ 2009, \mnras, 394, 577

\bibitem[{{Overzier} {et~al.}(2006){Overzier}, {Bouwens}, {Illingworth}, \&
  {Franx}}]{overzier06}
{Overzier}, R.~A., {Bouwens}, R.~J., {Illingworth}, G.~D., \& {Franx}, M.\ 2006,
  \apjl, 648, L5

\bibitem[{{Rees}(1989)}]{rees89}{Rees}, M.~J. 1989, \mnras, 239, 1P

\bibitem[{{Romano-D{\'{\i}}az} {et~al.}(2006){Romano-D{\'{\i}}az},
  {Faltenbacher}, {Jones}, {Heller}, {Hoffman}, \& {Shlosman}}]{erd06}
{Romano-D{\'{\i}}az}, E., {Faltenbacher}, A., {Jones}, D., {Heller}, C.,
  {Hoffman}, Y., \& {Shlosman}, I. 2006, \apjl, 637, L93

\bibitem[{{Romano-D{\'{\i}}az} {et~al.}(2007){Romano-D{\'{\i}}az}, {Hoffman},
  {Heller}, {Faltenbacher}, {Jones}, \& {Shlosman}}]{erd07}
{Romano-D{\'{\i}}az}, E., {Hoffman}, Y., {Heller}, C., {Faltenbacher}, A.,
  {Jones}, D., \& {Shlosman}, I. 2007, \apj, 657, 56

\bibitem[{{Romano-D{\'{\i}}az} {et~al.}(2009){Romano-D{\'{\i}}az}, {Shlosman},
  {Heller}, \& {Hoffman}}]{erd09}
{Romano-D{\'{\i}}az}, E., {Shlosman}, I., {Heller}, C., \& {Hoffman}, Y. 2009,
  \apj, 702, 1250

\bibitem[{{Romano-D{\'{\i}}az} {et~al.}(2010){Romano-D{\'{\i}}az}, {Shlosman},
  {Heller}, \& {Hoffman}}]{erd-sbh}
---. 2010, \apj, 716, 1095

\bibitem[{{Romano-D{\'{\i}}az} {et~al.}(2008){Romano-D{\'{\i}}az}, {Shlosman},
  {Hoffman}, \& {Heller}}]{erd08}
{Romano-D{\'{\i}}az}, E., {Shlosman}, I., {Hoffman}, Y., \& {Heller}, C. 2008,
  \apjl, 685, L105

\bibitem[{{Shapiro} \& {Raga}(2001)}]{shapiro01}
{Shapiro}, P.~R. \& {Raga}, A.~C. 2001, in Revista Mexicana de Astronomia y
  Astrofisica Conference Series, Vol.~10, Revista Mexicana de Astronomia y
  Astrofisica Conference Series, ed. {J.~Cant{\'o} \& L.~F.~Rodr{\'{\i}}guez},
  109--114

\bibitem[{{Shaver} {et~al.}(1996){Shaver}, {Wall}, {Kellermann}, {Jackson}, \&
  {Hawkins}}]{shaver96}
{Shaver}, P.~A., {Wall}, J.~V., {Kellermann}, K.~I., {Jackson}, C.~A., \&
  {Hawkins}, M.~R.~S. 1996, \nat, 384, 439

\bibitem[{{Shlosman} \& {Phinney}(1989)}]{phinney89}
{Shlosman}, I. \& {Phinney}, E.~S. 1989, unpublished

\bibitem[{{Sijacki} {et~al.}(2009){Sijacki}, {Springel}, \&
  {Haehnelt}}]{sijacki09}
{Sijacki}, D., {Springel}, V., \& {Haehnelt}, M.~G. 2009, \mnras, 400, 100

\bibitem[{{Springel} {et~al.}(2005){Springel}, {White}, {Jenkins}, {Frenk},
  {Yoshida}, {Gao}, {Navarro}, {Thacker}, {Croton}, {Helly}, {Peacock}, {Cole},
  {Thomas}, {Couchman}, {Evrard}, {Colberg}, \& {Pearce}}]{millenium}
{Springel}, V., et al. 2005, \nat, 435, 629

\bibitem[{{Stark} {et~al.}(2009){Stark}, {Ellis}, {Bunker}, {Bundy}, {Targett},
  {Benson}, \& {Lacy}}]{stark09}
{Stark}, D.~P., {Ellis}, R.~S., {Bunker}, A., {Bundy}, K., {Targett}, T.,
  {Benson}, A., \& {Lacy}, M. 2009, \apj, 697, 1493

\bibitem[{{Stark} {et~al.}(2007){Stark}, {Loeb}, \& {Ellis}}]{stark07}
{Stark}, D.~P., {Loeb}, A., \& {Ellis}, R.~S. 2007, \apj, 668, 627

\bibitem[{{Steidel} {et~al.}(2003){Steidel}, {Adelberger}, {Shapley},
  {Pettini}, {Dickinson}, \& {Giavalisco}}]{steidel03}
{Steidel}, C.~C., {Adelberger}, K.~L., {Shapley}, A.~E., {Pettini}, M.,
  {Dickinson}, M., \& {Giavalisco}, M. 2003, \apj, 592, 728

\bibitem[{{Stevens} {et~al.}(2010){Stevens}, {Jarvis}, {Coppin}, {Page},
  {Greve}, {Carrera}, \& {Ivison}}]{stevens10}
{Stevens}, J.~A., {Jarvis}, M.~J., {Coppin}, K.~E.~K., {Page}, M.~J., {Greve},
  T.~R., {Carrera}, F.~J., \& {Ivison}, R.~J. 2010, \mnras, 627

\bibitem[{{Stiavelli} {et~al.}(2005){Stiavelli}, {Djorgovski}, {Pavlovsky},
  {Scarlata}, {Stern}, {Mahabal}, {Thompson}, {Dickinson}, {Panagia}, \&
  {Meylan}}]{stiavelli05}
{Stiavelli}, M., et al. 2005, \apjl, 622, L1

\bibitem[{{Tasitsiomi} {et~al.}(2004){Tasitsiomi}, {Kravtsov}, {Gottl{\"o}ber},
  \& {Klypin}}]{t04}
{Tasitsiomi}, A., {Kravtsov}, A.~V., {Gottl{\"o}ber}, S., \& {Klypin}, A.~A.
  2004, \apj, 607, 125

\bibitem[{{Trenti} {et~al.}(2008){Trenti}, {Santos}, \&
  {Stiavelli}}]{trenti08b}
{Trenti}, M., {Santos}, M.~R., \& {Stiavelli}, M. 2008, \apj, 687, 1

\bibitem[{{Trenti} \& {Stiavelli}(2007)}]{trenti07}
{Trenti}, M. \& {Stiavelli}, M. 2007, \apj, 667, 38

\bibitem[{{Trenti} \& {Stiavelli}(2008)}]{trenti08}
---. 2008, \apj, 676, 767

\bibitem[{{Trenti} {et~al.}(2010){Trenti}, {Stiavelli}, {Bouwens}, {Oesch},
  {Shull}, {Illingworth}, {Bradley}, \& {Carollo}}]{trenti10}
{Trenti}, M., et al. 2010, \apjl, 714,
  L202

\bibitem[Trenti et al.(2010)]{trenti10b} Trenti, M., Smith, 
B.~D., Hallman, E.~J., Skillman, S.~W., \& Shull, J.~M.\ 2010, \apj, 711, 1198 

\bibitem[Utsumi et al.(2010)]{utsumi10} Utsumi, Y., Goto, T., Kashikawa, N., 
Miyazaki, S., Komiyama, Y., Furusawa, H., \& Overzier, R.\ 2010, \apj, 721, 1680

\bibitem[{{van de Weygaert} \& {Bertschinger}(1996)}]{vdw96}
{van de Weygaert}, R. \& {Bertschinger}, E. 1996, \mnras, 281, 84

\bibitem[{{Venemans} {et~al.}(2003){Venemans}, {Kurk}, {Miley}, \&
  {R{\"o}ttgering}}]{venemans03}
{Venemans}, B.~P., {Kurk}, J.~D., {Miley}, G.~K., \& {R{\"o}ttgering}, H.~J.~A.
  2003, New Astronomy Reviews, 47, 353

\bibitem[{{Venemans} {et~al.}(2007){Venemans}, {R{\"o}ttgering}, {Miley}, {van
  Breugel}, {de Breuck}, {Kurk}, {Pentericci}, {Stanford}, {Overzier}, {Croft},
  \& {Ford}}]{venemas07}
{Venemans}, B.~P., et al. 2007, \aap, 461, 823

\bibitem[{{Walter} {et~al.}(2004)}]{walter04}
{Walter}, F., {Carilli}, C., {Bertoldi}, F., {Menten}, K., {Cox}, P., {Fan}, X. \& 
   {Strauss}, M.~A. 2004, \apj, 615, L17

\bibitem[{{Willott} {et~al.}(2005){Willott}, {Crampton}, {Hutchings},
  {Sawicki}, {Simard}, {Jarvis}, {McLure}, \& {Percival}}]{willott05}
{Willott}, C.~J., et al. 2005, in Growing
  Black Holes: Accretion in a Cosmological Context, ed. {A.~Merloni,
  S.~Nayakshin, \& R.~A.~Sunyaev}, 102--107

\bibitem[{{Willott} {et~al.}(2010){Willott}, {Delorme}, {Reyl{\'e}}, {Albert},
  {Bergeron}, {Crampton}, {Delfosse}, {Forveille}, {Hutchings}, {McLure},
  {Omont}, \& {Schade}}]{willott10}
{Willott}, C.~J., et al. 2010, \aj, 139, 906

\bibitem[{{Wyithe} \& {Loeb}(2003)}]{wyithe03}
{Wyithe}, J.~S.~B. \& {Loeb}, A. 2003, \apj, 586, 693

\bibitem[{{Wyithe} \& {Loeb}(2006)}]{wyithe06}
---. 2006, \nat, 441, 322

\bibitem[{{York} {et~al.}(2000){York}, {Adelman}, {Anderson}, {Anderson},
  {Annis}, {Bahcall}, {Bakken}, {Barkhouser}, {Bastian}, {Berman}, {Boroski},
  {Bracker}, {Briegel}, {Briggs}, {Brinkmann}, {Brunner}, {Burles}, {Carey},
  {Carr}, {Castander}, {Chen}, {Colestock}, {Connolly}, {Crocker}, {Csabai},
  {Czarapata}, {Davis}, {Doi}, {Dombeck}, {Eisenstein}, {Ellman}, {Elms},
  {Evans}, {Fan}, {Federwitz}, {Fiscelli}, {Friedman}, {Frieman}, {Fukugita},
  {Gillespie}, {Gunn}, {Gurbani}, {de Haas}, {Haldeman}, {Harris}, {Hayes},
  {Heckman}, {Hennessy}, {Hindsley}, {Holm}, {Holmgren}, {Huang}, {Hull},
  {Husby}, {Ichikawa}, {Ichikawa}, {Ivezi{\'c}}, {Kent}, {Kim}, {Kinney},
  {Klaene}, {Kleinman}, {Kleinman}, {Knapp}, {Korienek}, {Kron}, {Kunszt},
  {Lamb}, {Lee}, {Leger}, {Limmongkol}, {Lindenmeyer}, {Long}, {Loomis},
  {Loveday}, {Lucinio}, {Lupton}, {MacKinnon}, {Mannery}, {Mantsch}, {Margon},
  {McGehee}, {McKay}, {Meiksin}, {Merelli}, {Monet}, {Munn}, {Narayanan},
  {Nash}, {Neilsen}, {Neswold}, {Newberg}, {Nichol}, {Nicinski}, {Nonino},
  {Okada}, {Okamura}, {Ostriker}, {Owen}, {Pauls}, {Peoples}, {Peterson},
  {Petravick}, {Pier}, {Pope}, {Pordes}, {Prosapio}, {Rechenmacher}, {Quinn},
  {Richards}, {Richmond}, {Rivetta}, {Rockosi}, {Ruthmansdorfer}, {Sandford},
  {Schlegel}, {Schneider}, {Sekiguchi}, {Sergey}, {Shimasaku}, {Siegmund},
  {Smee}, {Smith}, {Snedden}, {Stone}, {Stoughton}, {Strauss}, {Stubbs},
  {SubbaRao}, {Szalay}, {Szapudi}, {Szokoly}, {Thakar}, {Tremonti}, {Tucker},
  {Uomoto}, {Vanden Berk}, {Vogeley}, {Waddell}, {Wang}, {Watanabe},
  {Weinberg}, {Yanny}, \& {Yasuda}}]{york00}
{York}, D.~G., et al. 2000, \aj, 120, 1579

\bibitem[{{Zaroubi} {et~al.}(1995){Zaroubi}, {Hoffman}, {Fisher}, \&
  {Lahav}}]{zaroubi95}
{Zaroubi}, S., {Hoffman}, Y., {Fisher}, K.~B., \& {Lahav}, O. 1995, \apj, 449,
  446

\bibitem[{{Zheng} {et~al.}(2006){Zheng}, {Overzier}, {Bouwens}, {White},
  {Ford}, {Ben{\'{\i}}tez}, {Blakeslee}, {Bradley}, {Jee}, {Martel}, {Mei},
  {Zirm}, {Illingworth}, {Clampin}, {Hartig}, {Ardila}, {Bartko}, {Broadhurst},
  {Brown}, {Burrows}, {Cheng}, {Cross}, {Demarco}, {Feldman}, {Franx},
  {Golimowski}, {Goto}, {Gronwall}, {Holden}, {Homeier}, {Infante}, {Kimble},
  {Krist}, {Lesser}, {Menanteau}, {Meurer}, {Miley}, {Motta}, {Postman},
  {Rosati}, {Sirianni}, {Sparks}, {Tran}, \& {Tsvetanov}}]{zheng06}
{Zheng}, W., et al. 2006, \apj, 640, 574

\end{thebibliography}


\appendix

\section{Constrained Realizations: New Formalism}
\label{sec:a1}

One of the key ingredients of the canonical model is that the
primordial perturbation field constitutes a random Gaussian
field. Hence the setting of initial conditions (ICs) for cosmological
simulations amounts to sampling such fields. Gaussian random fields
are uniquely determined by their power spectrum, and the assumed
statistical homogeneity of the universe, and hence of the perturbation
field, implies that Fourier modes of the field are uncorrelated. It
follows that an unconstrained realization a of Gaussian field can be
performed by sampling the independent Fourier modes from a normal
distribution.  Here an unconstrained realization of a Gaussian filed
means that the realization obeys no other constraint but the assumed
power spectrum.

A constrained realization (CR) of a Gaussian field is a random
realization of such a field constructed to obey a set of linear
constraints imposed on the field. The imposed constraints violates the
statistical homogeneity of the field, hence the random sampling of the
Fourier modes cannot be used to construct CRs. The \citet[][hereafter
HR]{hr91} provided the optimal algorithm for the construction of CRs
and it used here for setting the ICs \citep[see][]{zaroubi95,hoffman09}.

The aim is to construct random realizations of a Gaussian field,
defined by the WMAP5 power spectrum, constrained to have a local
density maximum on a mass scale $M_{\rm s}$ centered at the center of
the computational box and designed to collapse by a redshift $z_{\rm
  coll}$. The linear overdensity corresponding to the assumed redshift
of collapse is calculated by the non-linear top-hat model
\citep[\eg][and refs. therein]{lokas04}.  The field under
consideration is the linear fractional overdensity, $\delta(\br)$, and
the constraints corresponds to the $\delta$ field smoothed with a
kernel corresponding to $M_{\rm s}$, $\Delta_{\rm s}(\br)$.  A local
density maximum is defined by 10 constraints: the value of
$\Delta_{\rm s}$, the three components of its gradient and the 6
independent components of the Hessian of the field. The three first
derivatives are set to zero, and the second derivatives are set
according the theory of the Gaussian statistics of local density
maxima \citep{bbks}.  This can be fully integrated into the HR
algorithm, but it introduces a considerable computational complexity
\citep{vdw96}. Alternatively the local maximum can be constructed by
constraining the value of $\Delta_{\rm s}$ at the location of the peak
and on six points located along three orthogonal directions, at
distances of the order of the smoothing length corresponding to
$M_{\rm s}$. Given the three angles defining the three orthogonal axes
these constitute 10 constraints. The off-peak values can be calculated
by using the full Gaussian statistics machinery of \citet{bbks}.

Before the proceeding to the construction of CRs the smoothing kernel
needs to be specified.  Analytic considerations favor the use of a
top-hat filter. However, the sharp real space cut-off leads to the
Gibbs Phenomenon, the so-called in k-space, namely the 'ringing' of
corresponding fields in k-space. To avoid that a Gaussian filter is
adopted here. Namely,
\begin{equation}
\label{eq;gf}
\Delta_{\rm s}(\br) = {1 \over (2 \pi R{^2_{\rm s}})^{3/2}} \int \exp\biggl( -{\vert \br -
    \br^{\prime}\vert^2 \over 2 R{_{\rm s}^2}}\biggr) \delta(\br^{\prime}) 
      \mathrm{d}^3\br^{\prime}.
\end{equation}
The smoothing length is related to $M_{\rm s}$ by 
\begin{equation}
\label{eq:rs}
R_s= 0.64 \biggl({ 3 M_{\rm s} \over 4 \pi \bar{\rho} }  \biggr)^{1/3},
\end{equation}
where $ \bar{\rho}$ is the mean cosmological density \citep{bbks}.

Given a random realization $\tilde{\delta}(\br)$ and a set of constraints 
$\{\Delta_{\rm s}(\br_\alpha)_{\alpha=1, ..., 7}\}$, where $\br_\alpha$ is the 
location at which the $\alpha$ constraint is imposed, a CR is obtained by:
\begin{equation}
\label{eq:cr}
\delta^{\rm CR}(\br) = \tilde{\delta}(\br) + \xi_{\delta,\Delta_{\rm s}}(\br - \br_\alpha)
  \Big(\xi_{\Delta_{\rm s},\Delta_{\rm 
    s}}(\br_\alpha\br_\beta)+\sigma{^2_\alpha}\delta{^K_{\alpha,
    \beta}} 
     \Big)^{-1}\Big( \Delta_{\rm s}(\br_\beta) - \tilde{\Delta}_{\rm s}(\br_\beta) \Big),
\end{equation}
where $P(k)$ is the power spectrum.  The various terms that appear in
Eq.\ref{eq:cr} are defined as follows. The cross-correlation function
of the underlying and the smoothed $\delta$ fields is given by:
\begin{equation}
\label{eq:xidD}
\xi_{\delta,\Delta_{\rm s}}= \sqrt{2\over\pi}  \int \exp\biggl( -{k^2 R{_{\rm s}^2} \over 2}\biggr) 
    P(k) k^2  \mathrm{d} k.
\end{equation}
The auto-correlation function of the smoothed field is given by:
\begin{equation}
\label{eq:xiDD}
\xi_{\Delta_{\rm s},\Delta_{\rm s}}= \sqrt{2\over\pi}  \int \exp\big( -k^2 R{_s^2} \big) P(k) k^2 
    \mathrm{d} k
\end{equation}
The pseudo constraints $\tilde{\Delta}_{\rm s}(\br_\beta)$ are
obtained by convolving the random $\tilde{\delta}(\br)$ with the
Gaussian filter, evaluated at the position of the constraints.  The
$\sigma{^2_\alpha}\delta{^K_{\alpha,\beta}}$ term in the inverse of
the auto-correlation matrix adds the possibility of a $\sigma_\alpha$
'fuzziness', or uncertainty, in the value of the imposed constraint
$\Delta_{\rm s}(\br_\alpha)$. ($\delta{^K_{\alpha,\beta}}$ is the
Kronecker $\delta$.)

The mean density field around a given density maximum is determined by
the assumed power spectrum, the height of the maximum and the Hessian
matrix \citep{bbks}.  It can be shown that for high peaks the mean
density profile around the peak is closely approximated by the mean
field around a random field point. This property is used here to
calculate the values of the off-peak constraints. The mean field
around a filed point, constrained to obey just $\Delta_{\rm
  s}(\br=0)$, is given by the Wiener Filter estimator
\citep{zaroubi95,hoffman09}, which is easily obtained from
Eq.~\ref{eq:cr} by setting $\tilde{\delta}(\br)$ to zero. It follows
that the construction of a CR proceeds in two steps. First, a single
constraint is imposed and the mean field given the constraint is
constructed.  The values of the off-peak constraints are extracted
from the mean field. Then, having the full set of constraints at hand
the ensemble of CRs is constructed.

The present implementation of the CR algorithm improves on the earlier
utilization of the method employed by \citet{erd06,erd07,erd08}.
There DM halos were imposed by using one single constraint for a given
halo, i.e. halos have been treated as random field points as far as
the constraints are considered. The present implementation results in
better constrained halos, which faithfully obey the imposed (linear)
constraints. In particular the virial mass of the imposed halo is very
close to the imposed $M_{\rm s}$ and it is situated very close to the
position imposed by the constraints.

 \begin{table*}
  \begin{center}
    \begin{tabular}{cccccc}
      \hline
      \# & X [Mpc~$h^{-1}$] & Y [Mpc~$h^{-1}$] & Z [Mpc~$h^{-1}$] & $\Delta_{\rm s}$ & 
      $\sigma$\\ 
      \hline
       1  &    0.0   &        0.0   &  0.0   &     9.05	  &   0.0  \\
       2  &  -0.94 &        0.0   &   0.0   &    8.39   &   0.1  \\
       3  &   0.94 &        0.0   &   0.0   &    8.39   &   0.1  \\
       4  &   0.0    &      -0.94 &   0.0   &    8.39   &   0.1  \\
       5  &   0.0    &       0.94 &   0.0   &    8.39   &   0.1  \\
       6  &   0.0    &       0.0   &  -0.94 &    8.39   &   0.1  \\
       7  &   0.0    &       0.0   &   0.94 &    8.39   &   0.1  \\
    \hline
    \end{tabular}
    \caption{ The positions and values of the constraints used to
      construct the set the initial conditions of the simulations. }
   \label{tab:const}
  \end{center}
\end{table*}

The set of constraints used for a mass scale $M_{\rm s}=10
^{12}M_\odot h^{-1}$, where $h$ is Hubble's constant in units of
$100~{\rm km~s^{-1}~Mpc^{-1}}$, collapse redshift of $z_{\rm
  coll}=6.0$ and the assumed WMAP5 cosmology is presented in
Table~\ref{tab:const}.
For simplicity the realizations are evaluated at the present epoch,
$z=0$, and are then scaled down their initial values by the linear
growth factor of the WMAP5 cosmology.


\end{document}